\documentclass{nature}
\usepackage{amsfonts}
\usepackage{amsmath}
\usepackage{amsthm}
\usepackage{amscd}
\usepackage{amssymb}
\usepackage{subfigure}
\usepackage{amsxtra}
\usepackage{gensymb}
\usepackage{bm}           
\usepackage{bbm}
\usepackage{graphicx}
\usepackage{epstopdf}
\usepackage{color}
\usepackage{times}
\usepackage{mdframed}
\usepackage[colorlinks=true,citecolor=blue,urlcolor=black]{hyperref}

\def\bi#1\ei {\begin{itemize}#1\end{itemize}}
\def\bn#1\en {\begin{enumerate}#1\end{enumerate}}
\def\bea#1\eea {\begin{align}#1\end{align}}
\def\bean#1\eean {\begin{align*}#1\end{align*}}
\def\ben#1\een {\begin{equation*}#1\end{equation*}}
\def\be#1\ee {\begin{equation}#1\end{equation}}
\def\bes#1\ees {\begin{equation}\begin{split}#1\end{split}\end{equation}}
\def\bear#1\eear {\begin{eqnarray}#1\end{eqnarray}}
\def\bear#1\eear {\begin{eqnarray*}#1\end{eqnarray*}}

\newcommand{\beq}{\begin{equation}}
\newcommand{\eeq}{\end{equation}}

\newcommand{\ket}[1]{\ensuremath{\left|#1\right\rangle}}

\newcommand{\order}[1]{\mathcal{O}(#1)}

\newcommand{\rv}[1]{{\bf{#1}}}

\title{\bf Observation of quantum fingerprinting beating the classical limit}

\author{%
Jian-Yu Guan$^{1,2 \ast}$, Feihu Xu$^{3 \ast}$, Hua-Lei Yin$^{1,2 \ast}$, Yuan Li$^{1,2}$, Wei-Jun Zhang$^{4}$, Si-Jing Chen$^{4}$, Xiao-Yan Yang$^{4}$, Li Li$^{1,2}$, Li-Xing You$^{4}$, Teng-Yun Chen$^{1,2}$, Zhen Wang$^{4}$, Qiang Zhang$^{1,2,5}$, Jian-Wei Pan$^{1,2}$
}

\begin{document}

\maketitle

\noindent $^{\ast}$ These authors contributed equally to this work. \\

\begin{affiliations}
 \item Department of Modern Physics and National Laboratory for Physical Sciences at Microscale, Shanghai Branch, University of Science and Technology of China, Hefei, Anhui 230026, China
 \item CAS Center for Excellence and Synergetic Innovation Center in Quantum Information and Quantum Physics, Shanghai Branch,  University of Science and Technology of China, Hefei, Anhui 230026, China
 \item Research Laboratory of Electronics, Massachusetts Institute of Technology, 77 Massachusetts Avenue, Cambridge, Massachusetts 02139, USA
 \item State Key Laboratory of Functional Materials for Informatics, Shanghai Institute of Microsystem and Information Technology, Chinese Academy of Sciences, Shanghai 200050, China
 \item Jinan Institute of Quantum Technology, Jinan, Shandong, 250101, China
\end{affiliations}


\begin{abstract}
Quantum communication has historically been at the forefront of advancements, from fundamental tests of quantum physics to utilizing the quantum-mechanical properties of physical systems for practical applications\cite{gisin2007quantum,lo2014secure}. In the field of communication complexity\cite{Yao1979}, quantum communication allows the advantage of an exponential reduction in the information transmitted over classical communication to accomplish distributed computational tasks\cite{BrassardQCC,RevModPhys.82.665}. However, to date, demonstrating this advantage in a practical setting continues to be a central challenge\cite{RevModPhys.82.665,TrojekQCC, HornFPs,du2006experimental,xu2015experimental}. Here, we report an experimental demonstration of a quantum fingerprinting protocol that for the first time surpasses the ultimate classical limit to transmitted information. Ultra-low noise superconducting single-photon detectors and a stable fibre-based Sagnac interferometer are used to implement a quantum fingerprinting system that is capable of transmitting less information than the classical proven lower bound over 20 km standard telecom fibre for input sizes of up to two Gbits. The results pave the way for experimentally exploring the advanced features of quantum communication and open a new window of opportunity for research in communication complexity and testing the foundations of physics\cite{RevModPhys.82.665}.
\end{abstract}


Quantum-communication network\cite{qiu2014quantum} is believed to be the next-generation platform for remote information processing tasks. So far, however, only one protocol -- quantum key distribution (QKD)\cite{lo2014secure} -- has been widely investigated and deployed in commercial applications. The extension of the practically available quantum communication protocols beyond QKD in order to fully understand the potential of large-scale quantum communication networks is therefore highly important. Significant progress has been made in this direction\cite{berlin2011exp,ng2012exp,clarke2012experimental,lunghi2013exp,liu2014exp,pappa2014experimental}, but the rich class of quantum communication complexity (QCC) protocols\cite{Yao1979,BrassardQCC,RevModPhys.82.665} remains largely undemonstrated, except for a few proof-of-principle implementations\cite{TrojekQCC, HornFPs, du2006experimental,xu2015experimental}. The field of QCC explores quantum-mechanical properties in order to determine the minimum amount of information that must be transmitted to solve distributed computational tasks\cite{BrassardQCC}. It not only has many connections to the foundational issues of quantum mechanics\cite{RevModPhys.82.665,steane2000physicists}, but also has important applications for the design of communication systems, green communication techniques, computer circuits and data structures\cite{kushilevitz2006communication}. For instance, QCC essentially connects the foundational physics questions regarding nonlocality with those of communication complexity studied in theoretical computer science\cite{RevModPhys.82.665}.

Quantum fingerprinting, proposed by Buhrman, Cleve, Watrous and Wolf, is the most appealing protocol in QCC\cite{QuantumFingerprinting}. Specifically, the simultaneous message-passing model\cite{Yao1979} corresponds to the scenario where two parties, Alice and Bob, respectively receive inputs $x_{a},x_{b}\in\{0,1\}^n$ and send messages to a third party, Referee, who must determine whether $x_a$ equals $x_b$ or not, with a small error probability $\epsilon$. This model has two requirements: (i) Alice and Bob do \emph{not} have access to shared randomness; (ii) there is one-way communication to Referee \emph{only}. Alice and Bob can achieve their goal by sending \emph{fingerprints} of their original inputs that are much shorter than the original inputs. It has been shown that the optimal classical protocols require fingerprints of a length that is at least $\order{\sqrt{n}}$\cite{newman1996public,babai1997randomized}, while, using quantum communication, Alice and Bob need to send fingerprints of only $\order{\log n}$ qubits\cite{QuantumFingerprinting,massar2005quantum}. Therefore, when the goal is to reduce the transmitted information, quantum communication provides an exponential improvement over the classical case. Despite this advantage, demonstrating it in a practical setting continues to be a challenge\cite{RevModPhys.82.665}.

Recently, a coherent-state quantum fingerprinting protocol for the realization using linear optics was proposed by Arrazola and L\"{u}tkenhaus\cite{arrazolaqfp}. On the basis of this protocol, Xu \emph{et al.} reported a proof-of-concept implementation that transmits less information than the best known classical protocol\cite{xu2015experimental}. Nonetheless, as noted already in ref.\cite{xu2015experimental}, a remaining question is ``whether quantum fingerprinting can beat the classical theoretical limit of transmitted information." This limit has been proven to be roughly two orders of magnitude smaller than the best known classical protocol\cite{babai1997randomized}, and surpassing it has been a long-standing experimental challenge. In this work, a quantum fingerprinting system is designed and demonstrated that for the first time beats the classical limit to transmitted information by up to 84\%.

\begin{figure*}
\includegraphics[width=1\columnwidth]{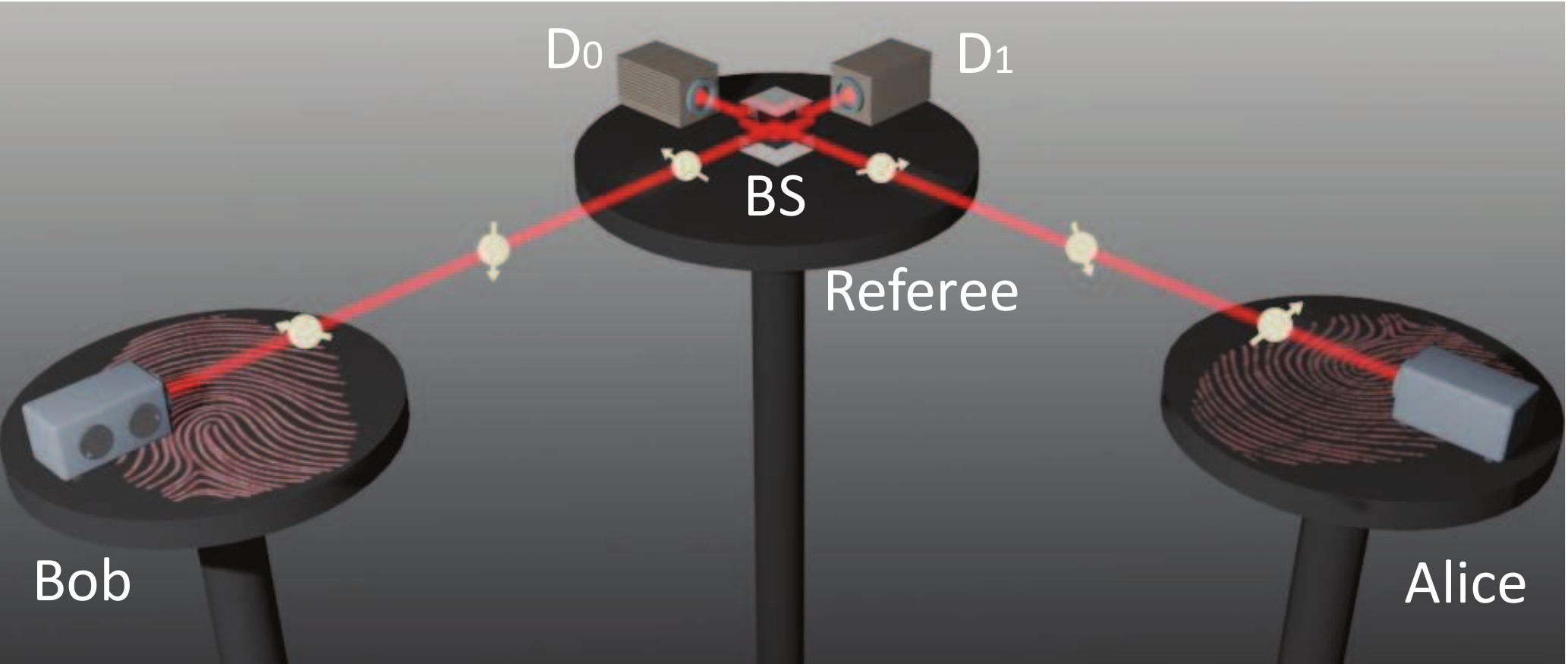}
\caption{(Colour online) A schematic illustration of the coherent-state quantum fingerprinting protocol. Alice and Bob use their input digital bits to modulate the phases of a sequence of weak coherent pulses and they send the sequence to Referee over two quantum channels. The incoming signals interfere at a beam-splitter (BS), and photons are detected in the output by two detectors $D_0$ and $D_1$. }\label{Fig:diagram}
\end{figure*}

As illustrated in Fig.~\ref{Fig:diagram}, the experiment adopted the coherent-state quantum fingerprinting protocol\cite{arrazolaqfp}. The detailed description of the protocol is presented in Table~\ref{table:protocol}. It has been proven that the quantum information $Q$ that can be transmitted by sending the sequence of weak coherent states satisfies\cite{arrazolaqfp}
\beq\label{Eq:scaling}
Q=O(\mu \log_2 n),
\eeq
where $\mu$ is defined as the total mean photon number in the entire pulse sequence. An important feature of the protocol is to fix $\mu$ to a small constant, which restricts the transmitted information (see Methods for details). For a fixed $\mu$, $Q$ corresponds to an exponential improvement over the classical case of $\order{\sqrt n}$ bits\cite{newman1996public,babai1997randomized}, which precisely provides the quantum advantage.

\begin{figure*}
  \centering
  \includegraphics[width=1\textwidth]{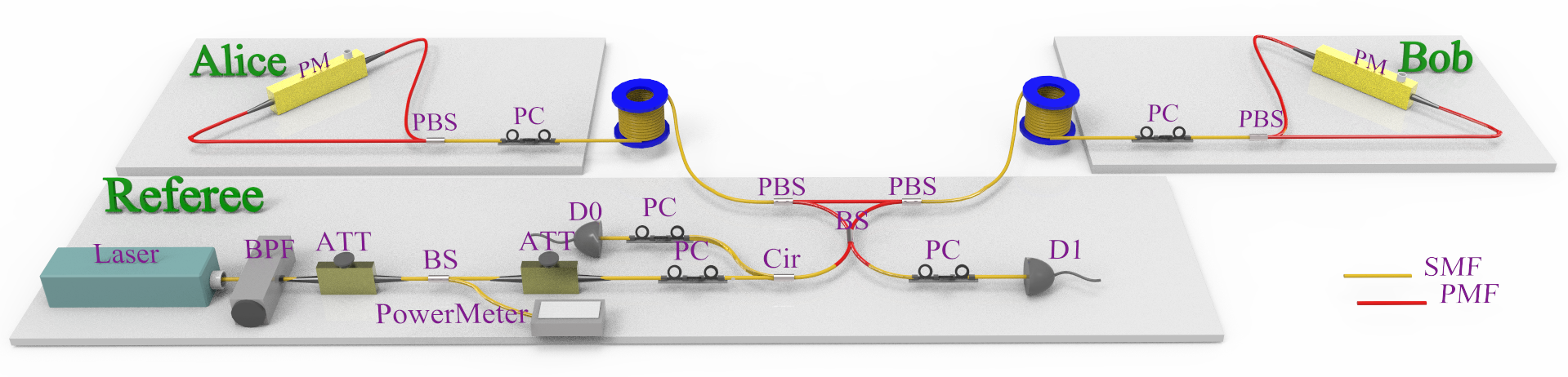}
  \caption{Experimental set-up of the quantum fingerprinting. Referee sends weak coherent pulses to Alice and Bob, who encode the phase of each of the pulses, using their phase modulator (PM) according to their codewords. The encoded pulses return and arrive simultaneously at Referee's input beam splitter (BS), where they interfere and are finally detected by two superconducting single-photon detectors ($D_0$ and $D_1$). See Methods for the experimental details. BPF: bandpass filter. ATT: attenuator. PC: polarization controller. Cir: circulator. PBS: polarization beam splitter. PMF: polarization maintaining fibre. SMF: standard single mode fibre.} \label{Fig:FPsetting}
\end{figure*}

To implement the coherent-state quantum fingerprinting protocol, the experiment utilizes a fibre-based Sagnac-type interferometer, as sketched in Fig.~\ref{Fig:FPsetting} (see Methods for the details). In this set-up, since the two pulses, sent from Referee to Alice and Bob, travel exactly the same path in the interferometer, two remarkable features are automatic compensation of the phase differences between the two pulses and high interference visibility. In experiment, one challenge is that Alice (Bob) should ensure that her (his) phase modulator (PM) modulate only the signal pulse, i.e., the one that returns from Bob (Alice), instead of the compensation pulse, i.e., the pulse that is sent directly from Referee. To do so, specific lengths of fibres and electrical cables are designed to separate the signal pulse from compensation pulse with 20 ns difference, and to carefully control the electrical gating signals applied to the PMs. Another challenge is that the coherent-state quantum fingerprinting protocol\cite{arrazolaqfp} requires the operation of the system at an ultra-low mean photon number, which is well below $10^{-7}$ per pulse. Indeed, as can be deduced from Eq.~\eqref{Eq:scaling}, a lower mean photon number leads to a reduction in the transmitted information, which permits the demonstration of beating the classical limit. To properly detect such a weak signal, advanced SNSPDs with on-chip narrow-band-pass filters\cite{yang2014superconducting, yang2015temperature} are installed. These SNSPDs have an ultra-low dark count rate of about 0.11 Hz and a high quantum efficiency of 45.6\% at 1532nm wavelength.

\begin{figure*}
  \centering
  \includegraphics[width=16cm]{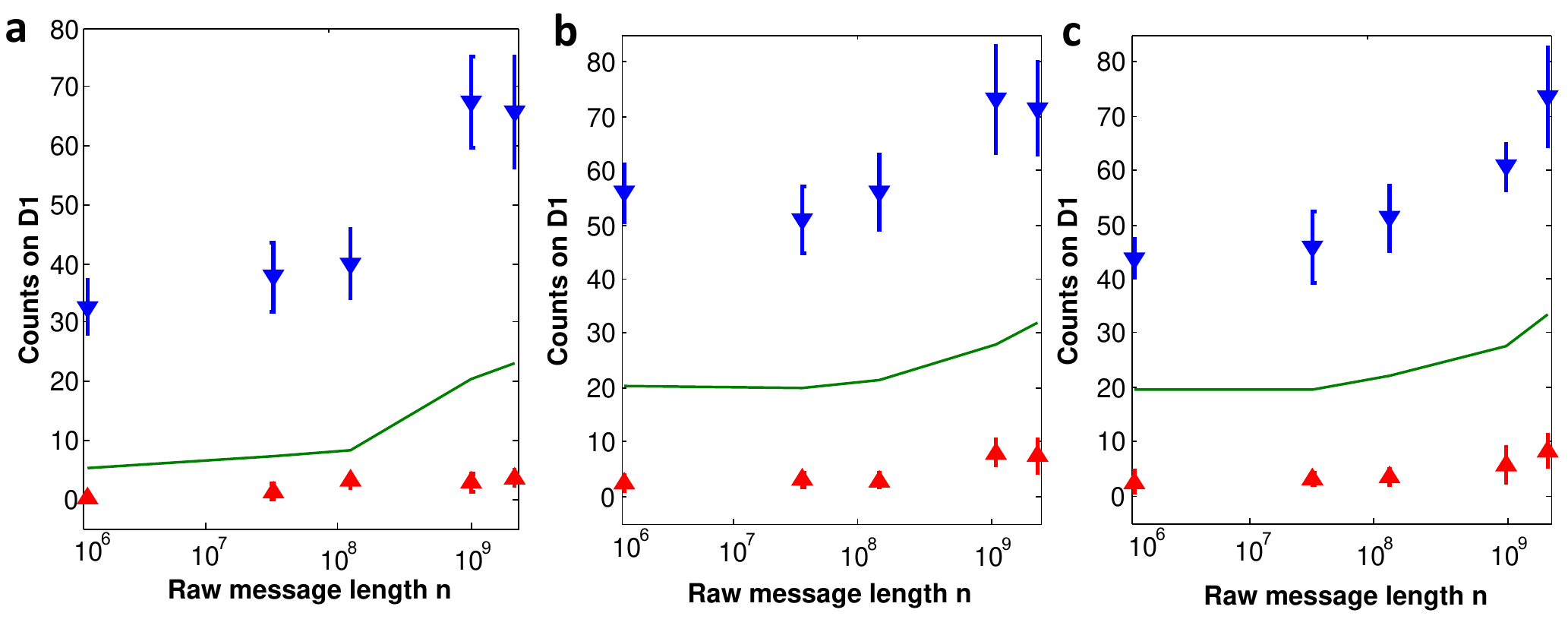}
  \caption{The experimental counts on $D_1$ for \textbf{a,} 0 km, \textbf{b,} 10 km, and \textbf{c,} 20 km. The blue points indicate the counts for two messages with $\delta=0.22$ difference, while the red points show the counts for two identical messages. The green curve is the threshold value $D_{1,th}$. The error bars correspond to one standard deviation, which is quantified by repeating the experiment ten times. }\label{Fig:counts}
\end{figure*}

To surpass the classical limit, the losses should be carefully controlled. The overall transmittance of Referee's PBS and BS is 80.16\% (78.5\%) from Alice (Bob) to Referee. The system is implemented with total distances (from Alice to Bob) of 0 km, 10 km and 20 km fibre spools, whose losses are characterised to be about 0 dB, 1.86 dB and 3.92 dB respectively. The distances between Alice (or Bob) and the referee are 0 km, 5 km and 10 km. Under each distance, five different message sizes $n$ are chosen as $2\times10^6$, $4\times10^7$, $1.42\times10^8$, $1\times10^9$, and $2\times10^9$. For each message, an ECC is applied based on the Toeplitz-matrices random linear code\cite{xu2015experimental}, which has a rate of $R=0.24$ and a minimum distance of $\delta=0.22$. The random numbers to construct the matrices are generated from a quantum random number generator\cite{xu2012ultrafast}. Here for a message up to two Gbits, the system requires several hours to transmit the pulses. Hence, a stable interference is essential for the experiment. The stability of the system is monitored, and the result is that, during 24 hours of continuous operation, the overall intensity fluctuations are less than 3.7\% and the interference visibility remains over 96\%. In the experiment, the key observation parameter is the number of counts on detector $D_1$. These experimental results are shown in Fig.~\ref{Fig:counts}. The clear difference between the worst-case different inputs with $\delta=0.22$ difference (blue points) and the identical inputs (red points) makes it possible to run the protocol (see Table~\ref{table:protocol}). In all the runs of experiments, a maximal error probability of $\epsilon=2.6\times10^{-5}$ (See Supplementary) was achieved. The maximum error probability was calculated from the theoretical model of the experiment\cite{arrazolaqfp,xu2015experimental}.

\begin{figure*}
  \centering
  \includegraphics[width=16cm]{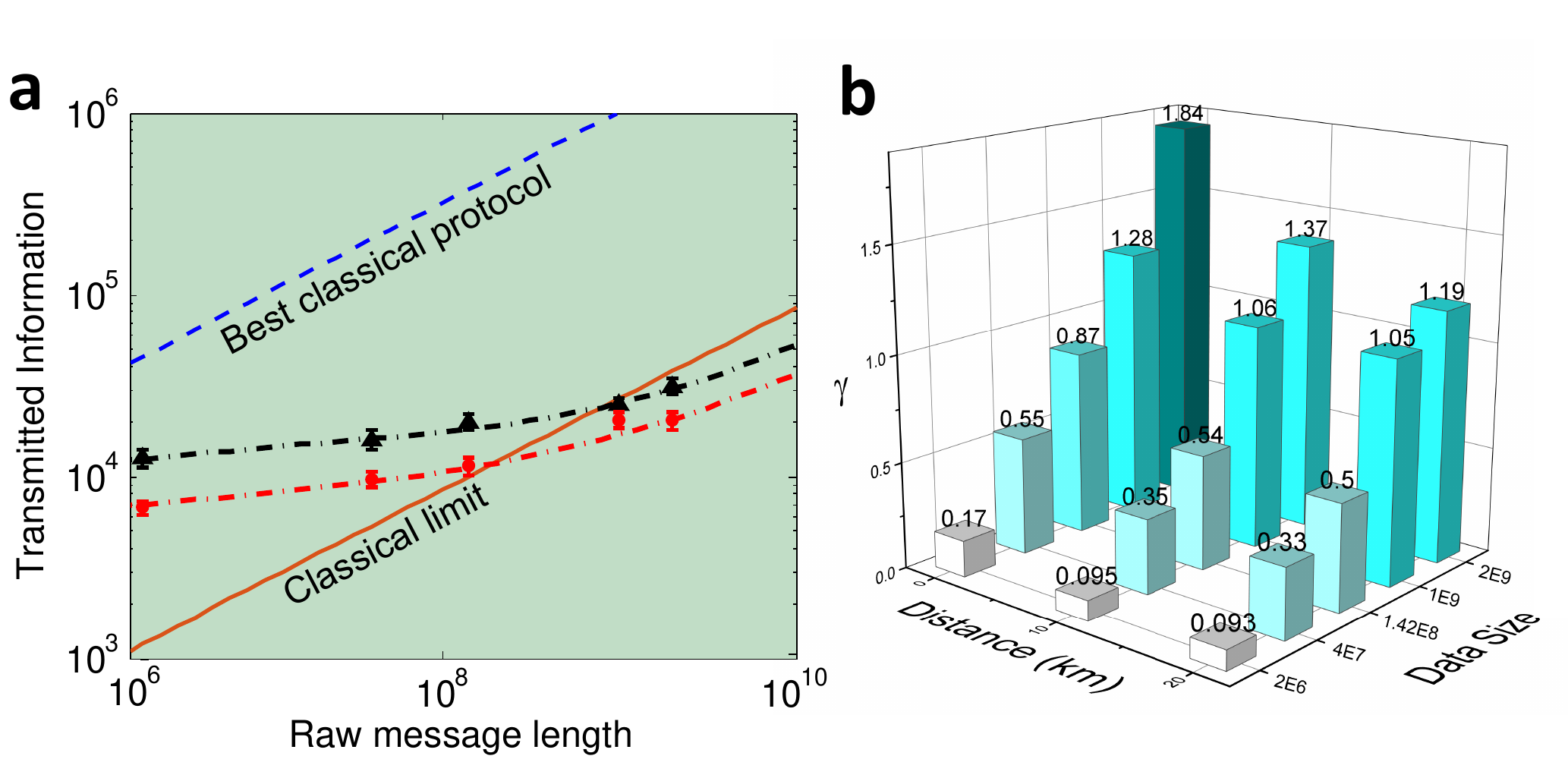}
  \caption{(Color online) \textbf{a,} Log-log plot of the total transmitted information. The red and black points are the experimental results at 0 km and 20 km respectively. For various $n$, the transmitted information of our experimental quantum fingerprinting protocol is much lower than the transmitted information of the best known classical algorithm. For large $n$, our results are, in strict terms, better than the classical limit for a wide range of practical values of the input size. \textbf{b,} The ratio $\gamma$ between classical limit $C_{\text{limit}}$ and the transmitted quantum information $Q$. For the three small input sizes, no advantage over the classical limit was obtained. However, for the two large input sizes, the ratio is well above one over different fibre distances. Our experiment transmitted as much as 84\% less information than the classical limit.
  }\label{Fig:Information}
\end{figure*}

Fig.~\ref{Fig:Information}a shows the experimental transmitted information at 0 km (red data points) and 20 km (black data points) for different message sizes. The error bars come from the uncertainty in the estimation of the mean photon number $\mu$. In this figure, our quantum fingerprinting is compared with the classical limit (solid-orange curve) and the best known classical protocol (dashed-blue curve). The best known classical protocol needs to transmit at least $32\sqrt{n}$ bits of information\cite{babai1997randomized}. On the basis of the references\cite{newman1996public,babai1997randomized}, we prove an optimized bound for the classical limit (see Supplementary)\footnote{After completing our manuscript, we become aware that a similar optimized classical lower bound has been proved independently by Dave Touchette, Juan Miguel Arrazola and Norbert L\"{u}tkenhaus, who will post their results soon.}. This bound is given by
\beq\label{Eq:classicallimit}
C_{\text{limit}}=(1-2\sqrt{\epsilon})\sqrt{\frac{n}{2\ln2}}-1.
\eeq
Fig.~\ref{Fig:Information}a indicates that, with the increase of input size $n$, the classical limit scales linearly in the log-log plot, while the transmitted quantum information remains almost a constant. The transmitted information is up to two orders of magnitude lower than that in the previous experiment\cite{xu2015experimental}. Importantly, for large $n$, these experimental results clearly beat the classical limit for a wide range of practical values of the input size. To further illustrate our results, $\gamma$ is defined as the ratio between the classical limit $C_{\text{limit}}$ and the transmitted quantum information $Q$, i.e., $\gamma=C_{\text{limit}}/Q$. A value $\gamma>1$ implies that the classical limit is surpassed by our quantum fingerprinting protocol. In Fig.~\ref{Fig:Information}b, $\gamma$ is plotted as a function of different fibre distances and input data sizes. For the input sizes larger than one Gbit, $\gamma$ is well above one. The ratio is as large as $\gamma=1.84$, which implies that our quantum fingerprinting implementation beats the classical limit by up to 84\%.

Finally, to show the ability of the quantum protocol in the real world, two video files with sizes of two Gbits\cite{Video} were experimentally fingerprinted over 20 km fibre. A 14\% reduction in the transmitted information was obtained, as compared to the classical limit (see Supplementary Table 4 for details).


To conclude, by using ultra-low dark count superconducting detectors (i.e., $\sim 0.1$ Hz), as well as an automatic-phase compensation Sagnac system, a quantum-enhanced method for fingerprinting to beat the ultimate classical theoretical limit was demonstrated. With $\sim$1300 transmitted photons as the information carrier, two $\sim$2 Gbits video files were experimentally fingerprinted over 20 km fibres, and the potential for practical applications was thus indicated. In the future, our system can be deployed in the metropolitan fibre network for a field test. Higher detection efficiency SNSPDs\cite{superconducting-93} can be exploited to increase the transmission distance and reduce the required message length. Alice and Bob can also hold independent laser sources and apply the phase-locking techniques to interfere the weak coherent pulses. Last but not least, we point that many connections between communication complexity and foundational issues of quantum mechanics have been recently discovered\cite{RevModPhys.82.665}. In particular, quantum communication complexity is intimately linked to quantum nonlocality. Our experiment provides a first step in the development of experimental quantum communication complexity, which could even lead new proposals for experiments that test the foundations of physics.


\begin{methods}
\textbf{Transmitted information.}
In the coherent-state quantum fingerprinting protocol, the sequence of coherent states prepared by Alice and Bob are given by\cite{arrazolaqfp}
\begin{eqnarray}\label{coherentfpstates}
\ket{\phi_a} &=& \bigotimes_{j=1}^m\ket{e^{i[\pi E(x_a)_j+\theta_j]}\frac{\alpha}{\sqrt{m}}}_j, \\ \nonumber
\ket{\phi_b} &=& \bigotimes_{j=1}^m\ket{e^{i[\pi E(x_b)_j+\theta_j]}\frac{\alpha}{\sqrt{m}}}_j.
\end{eqnarray}
Here $E(x_a)_j$ ($E(x_b)_j$) is the $j$th binary bit of Alice's (Bob's) codeword, $\alpha$ is a complex amplitude, and $\theta_j$ -- which is assumed to be the same for Alice and Bob -- is the global phase of $j$th pulse. We define $\mu:=|\alpha|^2$ as the total mean photon number in the entire pulse sequence. The states of Eq.~\eqref{coherentfpstates} can be understood as a coherent-state version of an encoding of $m$-dimensional quantum states into states of a single photon across $m$ modes\cite{arrazolaqfp}. Importantly, by fixing $\mu$ to a small constant, we are restricting ourselves to an exponentially small subspace of the larger Hilbert space associated with the optical modes. This in turn restricts the capability of these systems to transmit information. Thus, to achieve the central goal of a reduction in the transmitted information, our protocol must use a number of modes that is linear in the input size $n$, with the significant advantage that the total mean photon number $\mu$ is very small. In particular, the number of photons used is more than quadratically smaller than in a classical protocol using photonic bits, where $O(\sqrt{n})$ photons are needed compared to $O(1)$ photons in the quantum case. This quantum advantage essentially provides an exponential improvement over the classical case.

\textbf{Experimental details.}
The laser source at Referee emits weak coherent pulses with 300 ps pulse-width at 1532 nm, which are filtered by a tunable band-pass filter (BPF) to suppress the spontaneous emission noise. The repetition rate of the pulsed laser is 25 MHz. In this rate, the spontaneous emission noise is about 10 dB lower than the signal, while this noise is suppressed down to 40 dB by the BPF. The pulses are separated by a 50/50 beam splitter (BS) into two pulses: one pulse is measured by a power meter to monitor the light intensity and the other one -- which is called the encoding pulse -- is injected into the interferometer. The encoding pulse is first attenuated and separated again at a polarization-maintaining BS into \emph{left pulse} and \emph{right pulse}. Note that this polarization-maintaining BS allows \emph{only} the light with slow-axis polarization to pass through. We define this polarization as $\rv{H}$. The left pulse passes through a polarization beam splitter (PBS) and travels to Alice over a fiber spool. Alice uses a polarization controller (PC) to compensate the polarization rotation (or drift) of the fiber spool that connects to her. After the polarization compensation, the pulse travels out from the $\rv{H}$ port of Alice's PBS, passes through a phase modulator (PM) \emph{without} any modulations and returns to the PBS from its $\rv{V}$ port. When the pulse returns to Referee, the polarization of the pulse has been rotated to $\rv{V}$. Hence the pulse passes through the $\rv{V}$ port of referee's PBS and travels to Bob. Note that the pulse travelling from Alice to Bob does not contain any information about Alice's codeword. This guarantees that there is no communication between Alice and Bob. Bob uses his PC to compensate the polarization rotation of the fiber spool that connects from the referee to him. Once the pulse passes through Bob's PM from his PBS's $\rv{V}$ port, Bob modulates the phase of the pulse according to his codeword $E(x_b)$. Similarly, the right pulse from the referee travels to Bob for a polarization compensation first and then goes to Alice for encoding according to Alice's codeword $E(x_a)$. Finally, the left pulse and the right pulse return simultaneously at Referee's polarization maintaining BS, where they interfere and are detected by two superconducting nanowire single photon detectors (SNSPDs). In front of the SNSPD, PC is used to optimize the detection efficiency. A time-to-digital converter records all the detected events and a 2.5 ns effective time window is used to detect the pulses. The electric signals that control the optical devices are generated from a AWG70002 waveform generator (manufactured by Tektronix).

\end{methods}



\begin{addendum}
 \item The authors thank J. M. Arrazola, N. L\"{u}tkenhaus, H.-K. Lo, X. Xie, M. Jiang for valuable discussions. Particularly, we thank Mike W. Wang for his help on the plot of Figure 1 and the implementation of ECC. This work was supported by the National Fundamental Research Program (under Grant No. 2011CB921300 and 2013CB336800), the National Natural Science Foundation of China, the Chinese Academy of Science, the 10000-Plan of Shandong Province. And F. Xu acknowledges the support from the Office of Naval Research (ONR) and the Air Force Office of Scientific Research (AFOSR).
\end{addendum}

\newpage

\begin{table*}
\begin{center}
\begin{tabular}{p{0.95\linewidth}}\hline \hline 
\begin{description}
\begin{footnotesize}
\item[1. Preparation.] Alice applies an error-correcting code (ECC) to her input $x_{a}$ of $n$ bits and generates a codeword $E(x_a)$ of $m=n/R$ bits, with $R$ indicating the rate of ECC. Then she prepares a sequence of $m$ weak coherent pulses and uses the codeword to modulate the phase of each pulse. The sequence of coherent states can be understood as a coherent version of the encoding of a single photon across $m$ modes. Bob completes a process that is the same as Alice's for his input $x_{b}$.

\item[2. Distribution.] Both Alice and Bob send their pulse trains to the Referee over two quantum channels. By using a phase interferometer, Referee interferes the individual pulses in a balanced beam-splitter (BS) and observes the clicks at the outputs of the BS, using two single-photon detectors, which are labelled ``$D_0$" and ``$D_1$". This process allows Referee to verify whether the relative phases of the incoming pulses are same or different\cite{andersson2006experimentally}. In an ideal situation, a click in detector $D_1$ will never happen if the phases of the incoming states are equal.

\item[3. Decision.]In the presence of experimental imperfections such as detector dark counts and imperfect interference, detector $D_1$ may fire even when the inputs are equal. However, in a case of small imperfections, the total number of clicks on $D_1$ for different inputs is much larger than the total number of clicks for equal inputs. A decision rule for Referee is employed on the basis of only the total number of clicks observed in detector $D_1$\cite{xu2015experimental}. Referee sets a threshold value $D_{1,th}$ such that, if the number of clicks is smaller than or equal to $D_{1,th}$, he will conclude that the inputs are equal. Otherwise, he concludes that they are different. In the protocol, the value of $D_{1,th}$ is chosen in such a way that an error is equally likely to occur in both cases.
\end{footnotesize}
\end{description}\\
\hline \hline
\end{tabular}
\end{center}
\caption{A detailed description of the coherent-state quantum fingerprinting protocol.} \label{table:protocol}
\end{table*}

\end{document}